\documentclass[iop]{emulateapj}

\usepackage {epsfig,graphicx,color}
\usepackage {txfonts   }
\usepackage {natbib    }
\usepackage {float     }
\usepackage {enumerate }
\usepackage {tabularx  }
\usepackage {url       }
\usepackage {capt-of   }
\usepackage {color     }

\begin{document}

%
%

\title{Cloudless atmospheres for L/T dwarfs and extra-solar
giant planets}

\author{ 
P. Tremblin\altaffilmark{1,2}      and
D. S. Amundsen\altaffilmark{1,3,4} and 
G. Chabrier\altaffilmark{1,5}      and
I. Baraffe\altaffilmark{1,5}       and
B. Drummond\altaffilmark{1}      and
S. Hinkley\altaffilmark{1}         and
P. Mourier\altaffilmark{5,6}       and
O. Venot\altaffilmark{7}
       }

\altaffiltext{1}{
  Astrophysics Group, University of Exeter, EX4 4QL Exeter, UK}

\altaffiltext{2}{
  Maison de la Simulation, CEA-CNRS-INRIA-UPS-UVSQ, USR 3441, Centre
  d'\'etude de Saclay, 91191 Gif-Sur-Yvette, France}

\altaffiltext{3}{Department of Applied Physics and Applied
  Mathematics, Columbia University, New York, NY 10025, USA}

\altaffiltext{4}{NASA Goddard Institute for Space Studies, New York,
  NY 10025, USA} 

\altaffiltext{5}{
  Ecole Normale Sup\'erieure de Lyon, CRAL, UMR CNRS 5574, 69364 Lyon
  Cedex 07, France}

\altaffiltext{6}{D\'epartement de Physique, Ecole Normale
  Sup\'erieure, 24 Rue Lhomond, F-75005 Paris, France} 

\altaffiltext{7}{
  Instituut voor Sterrenkunde, Katholieke Universiteit Leuven,
  Celestijnenlaan 200D, 3001 Leuven, Belgium} 

\email{tremblin@astro.ex.ac.uk or pascal.tremblin@cea.fr}

\begin{abstract}
  The admitted, conventional scenario to explain the complex spectral
  evolution of brown dwarfs (BD) since their first detections twenty
  years ago, has always been the key role played by micron-size
  condensates, called ''dust'' or ''clouds'', in their atmosphere. This
  scenario, however, faces major problems, in particular the J-band
  brightening and the resurgence of FeH absorption at the L to T
  transition, and a physical first-principle understanding of
    this transition is lacking. In this paper, we propose a new, completely
  different explanation for BD and extrasolar giant planet (EGP)
  spectral evolution, without the need to invoke clouds. We show that,
  due to the slowness of the CO/CH$_4$ and N$_2$/NH$_3$ chemical reactions,
  brown dwarf (L and T, respectively) and EGP atmospheres are subject
  to a thermo-chemical instability similar in nature to the fingering
  or chemical convective instability present in Earth oceans and at
  the Earth core/mantle boundary. The induced small-scale turbulent
  energy transport reduces the temperature gradient in the atmosphere,
  explaining the observed increase in near infrared J - H and J - K
  colors of L dwarfs and hot EGPs, while a warming up of the deep
  atmosphere along the L to T transition, as the CO/CH$_4$ instability vanishes,
  naturally solves the two aforementioned puzzles, and provides
    a physical explanation of the L to T transition. This new picture
  leads to a drastic revision of our understanding of BD and EGP
  atmospheres and their evolution.

\end{abstract}


\maketitle

%
%

\section{Introduction}\label{sect:intro}

The role of clouds in brown dwarf (BD) and extra-solar giant planet
(EGP) atmospheres has been an intense subject of research for the past
twenty years, since the first observations of these objects
\citep[e.g.][]{Tsuji:1996vu,Allard:2001fh,Ackerman:2001gka,Burrows:2006ia,Marley:2010kx,Morley:2014gs}. While cloudy atmosphere models reproduce the observed reddening
in infrared (IR) J - H and J - K colors
\citep{Burrows:2006ia,Saumon:2008im,Allard:2001fh}, several problems
remain unsolved:
\begin{itemize}
\item The driving mechanism of the cloud dynamics remains 
  poorly understood. Although convective overshooting has
  been proposed for such a mechanism \citep{Freytag:2010bh}, the low
  variability of L dwarfs \citep[][and references therein]{Metchev:2015dr}, which points to a relatively steady
  process, seems incompatible with the transient nature of
  overshooting. Furthermore,
  the change of regime in this mechanism and the sharpness in effective temperature of the LT
  transition, so in short the very physical nature of this transition
  remain  so far unexplained.
\item Neither the J band brightening nor the FeH resurgence at the L/T
  transition are understood. Although holes in the cloud
  cover can reproduce these 
  features \citep{Ackerman:2001gka,Burgasser:2002jq,Marley:2010kx},
   \citet{Buenzli:2015tn}
  recently showed that the variability of the spectrum is not
  compatible with holes but rather with height variations.
  The conclusion of \citet{Buenzli:2015tn} is indeed that the
  explanation for the re-emergence of FeH still remains to be found.
\item The absorption feature at 10 $\mu$m is not well reproduced
  with the current cloud models. Although some observations do suggest the
  presence of silicate dust \citep{Cushing:2006hx}, the absorption
  feature is not present in all observed spectra and, for the ones
  that display it \citep[e.g. 2M224, 2M0036, 2M1507, and 2M2244
    in][]{Stephens:2009cc}, cloud models cannot reproduce {\it both} the NIR
  reddening and the 10-$\mu$m absorption. 
\item Signatures of cloud polarization expected from cloud models
  remain undetected \citep{Goldman:2009jg}. Even though L dwarfs might not
  rotate fast enough nor have low enough surface gravity to be
  sufficiently oblate to produce detectable polarization \citep{Sengupta:2010ht}, this lack of detection questions the cloud hypothesis.
\end{itemize}

In \citet{Tremblin:2015fx}, we showed that the spectra and near-IR colors of T and Y
dwarfs can be reproduced without clouds if (i) non-equilibrium
chemistry of NH$_3$ is taken into account, and (ii) the temperature
gradient in the atmospheres of T dwarfs is reduced.
In this paper, we explore a similar mechanism
for L dwarfs and hot EGPs, using Denis-P 0255 and HR8799c,
respectively, as typical templates, and show that these effects
adequately reproduce both the FeH resurgence and the J-band
brightening at the L/T transition. We show that
a physical process responsible for the temperature gradient reduction in L and T dwarf
atmospheres is the chemical or (compositional) convection
  triggered by a thermo-chemical instability at the CO/CH$_4$ and N$_2$/NH$_3$
transitions.
Finally, we discuss the implications and the
limitations of this scenario. 

%
%
\begin{figure*}[t]
\centering
\includegraphics[width=\linewidth]{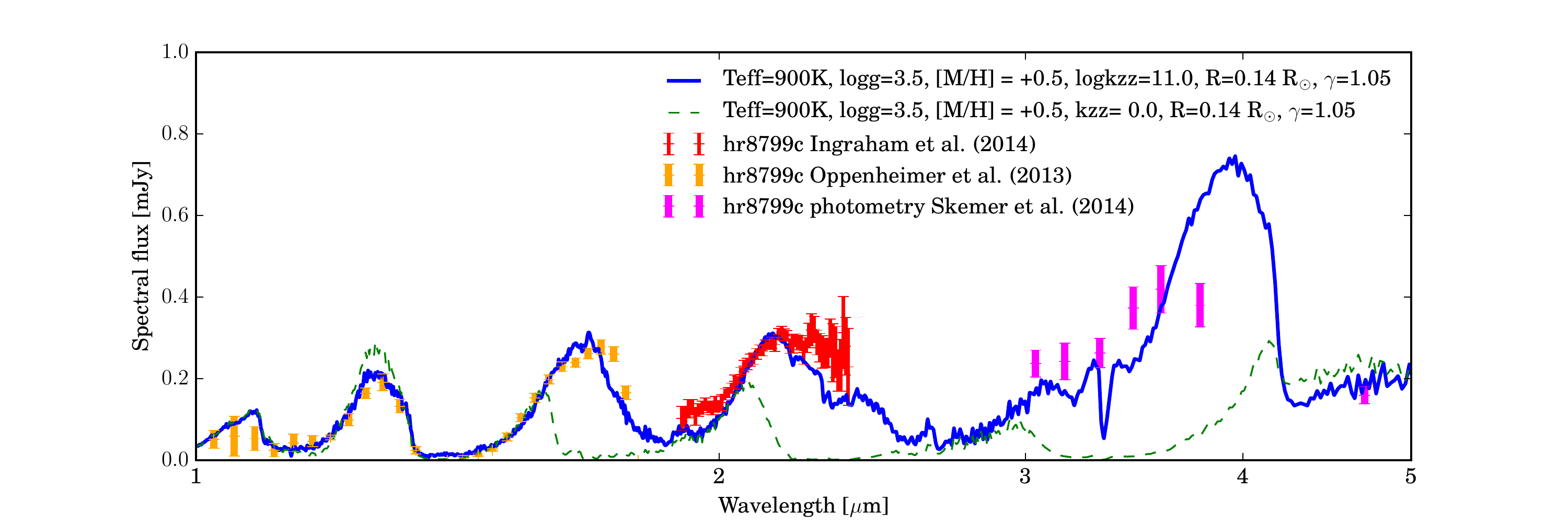}
\caption{\label{fig:spec_hr8799c} Spectral modeling of HR8799c. The
  blue spectrum is a model with out-of-equilibrium chemistry with
  $K_{zz}=10^{11}$~cm$^2$s$^{-1}$.  The green dashed model shows the
  corresponding equilibrium model with the same PT structure in order
  to illustrate the effect of the change in the CH$_4$ abundance profile.}
\end{figure*}

\begin{figure*}[t]
\centering
\includegraphics[width=\linewidth]{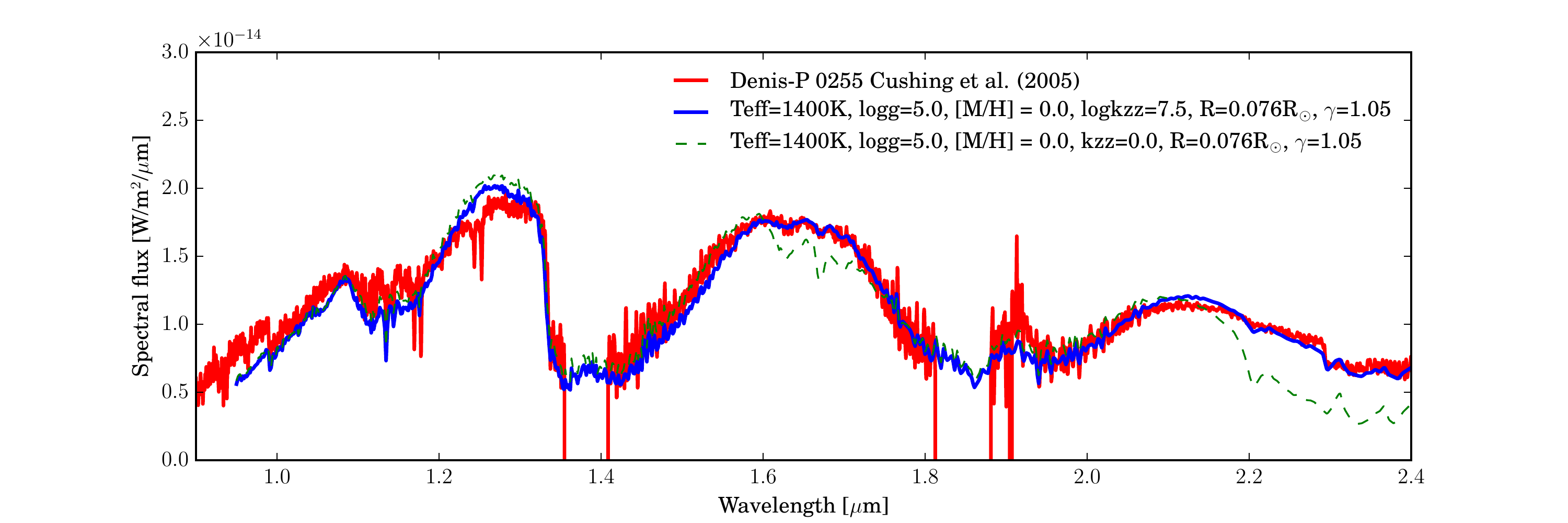}
\caption{\label{fig:spec_d0255} Spectral modeling of the L8 dwarf
  Denis-P 0255. The blue model is a model with out-of-equilibrium chemistry with
  $K_{zz}=10^{7.5}$~cm$^2$s$^{-1}$. The green dashed model shows the
  corresponding equilibrium model with the same PT structure in order
  to illustrate the effect of the change in the CH$_4$ abundance profile.}
\end{figure*}

\section{Spectral models for HR8799c and Denis-P
  0255}\label{sect:spec}

\begin{figure*}[t]
\centering
\includegraphics[width=0.49\linewidth]{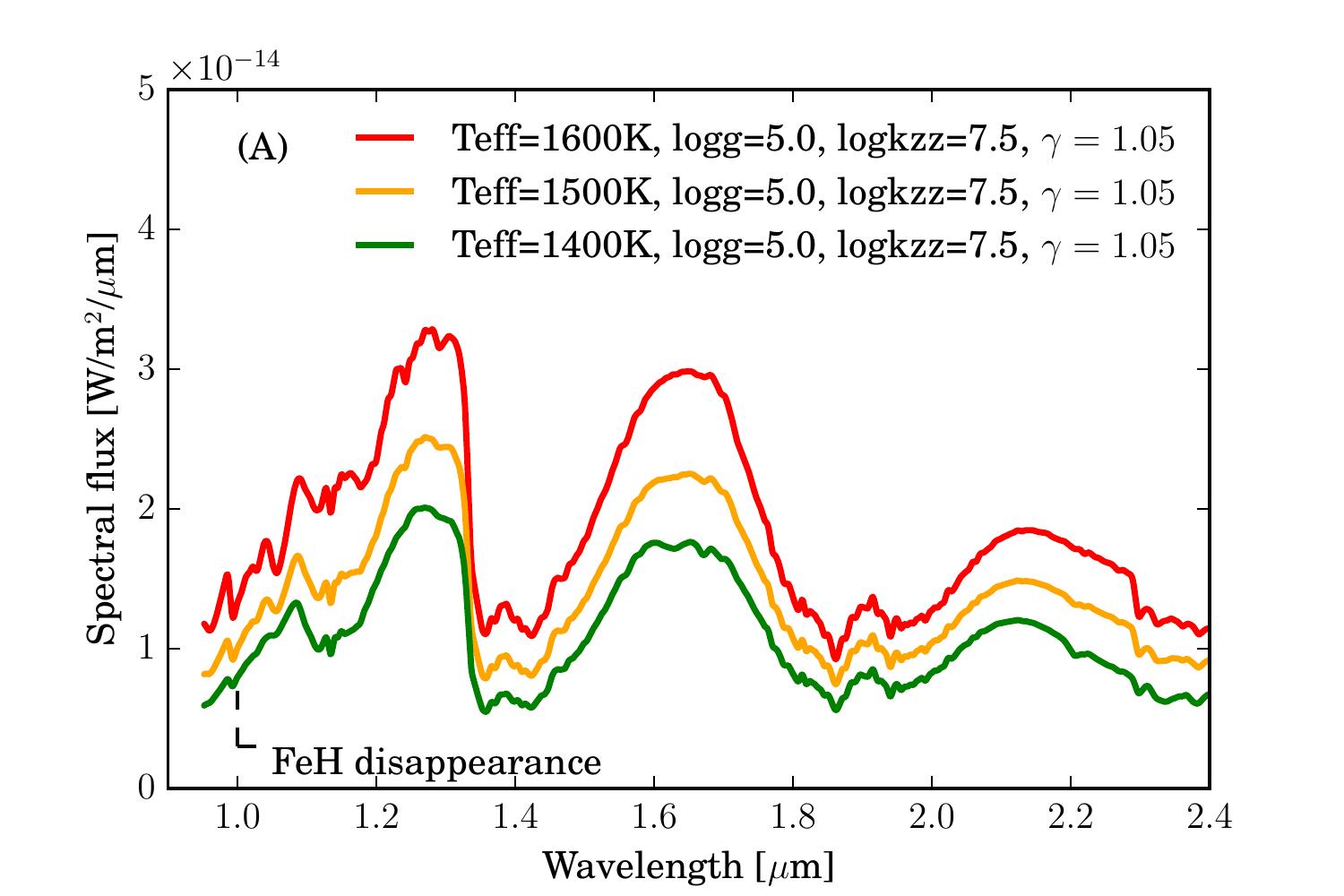}
\includegraphics[width=0.49\linewidth]{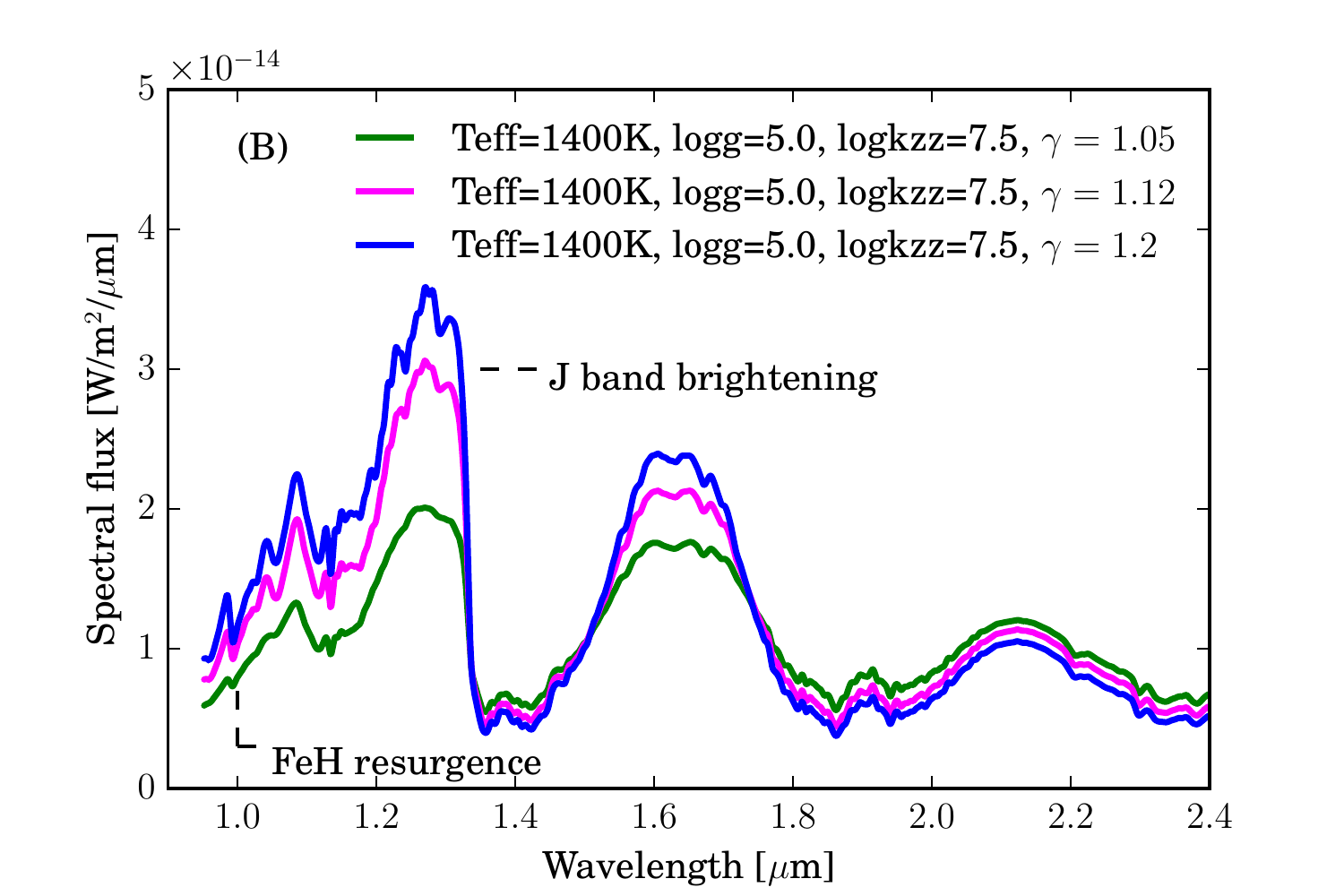}
\caption{\label{fig:spec_lt} Left: spectral sequence along L dwarfs showing the
  disappearance of FeH with lower effective temperature. Right:
  spectral sequence along the L/T transition at constant effective
  temperature, showing the resurgence of FeH and the brightening of
  the J band. }
\end{figure*}

\begin{figure}[t]
\centering
\includegraphics[width=\linewidth]{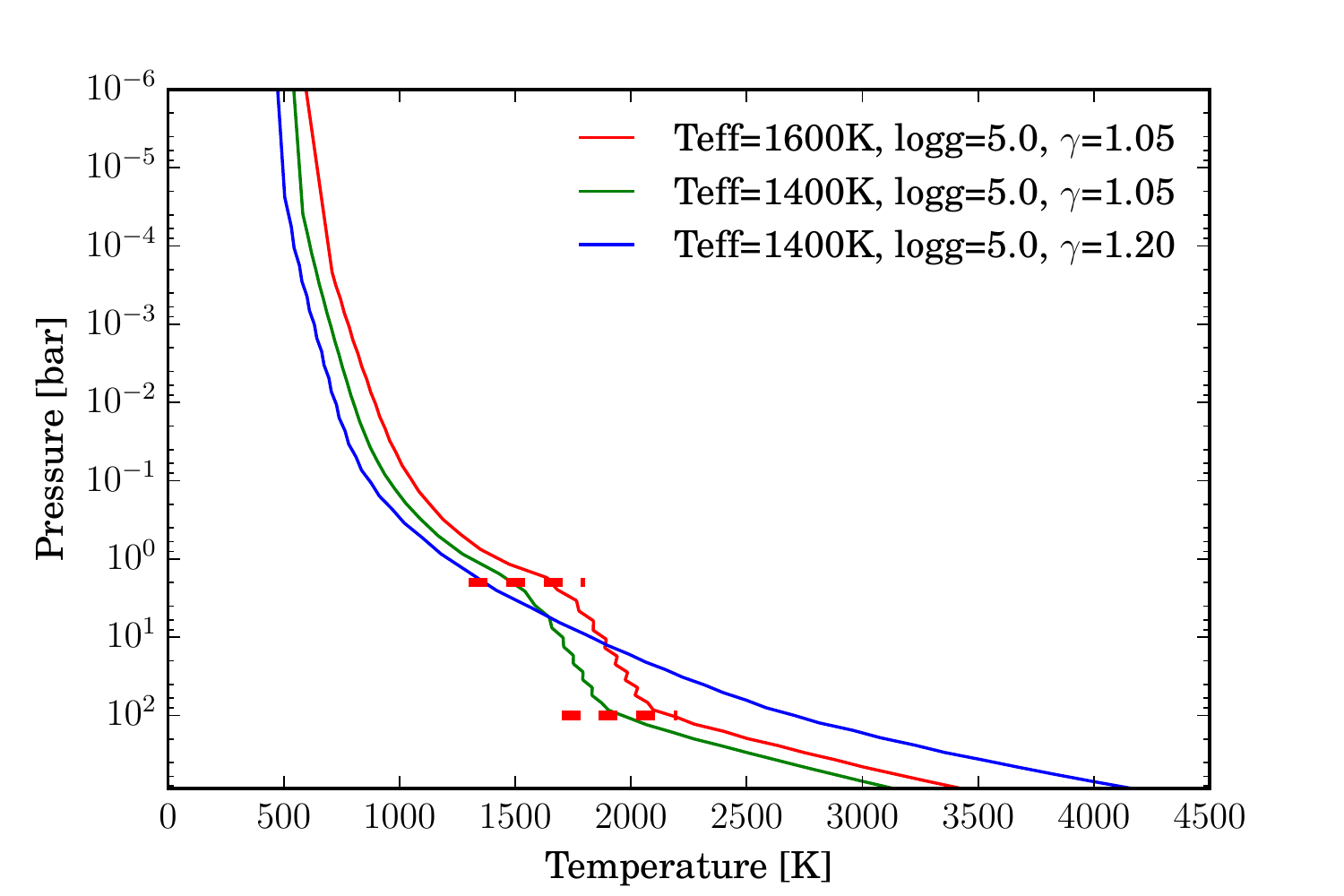}
\caption{\label{fig:pt} Pressure/Temperature structures of our models
  along the L sequence (T$_\mathrm{eff}$=1600K to T$_\mathrm{eff}$=1400K) and along the L/T
  transition at constant effective temperature ($\gamma$=1.05 to
  $\gamma$=1.2).
  }
\end{figure}

We use the same setup as described in \citet{Tremblin:2015fx} i.e., a 1D spectral code
ATMO with correlated-k coefficients for the radiative transfer,
developed and tested in \citet{Amundsen:2014df}, coupled to the CHNO-based chemical
network of \citet{Venot:2012fr}. We improved our chemical and opacity database with
the inclusion of the FeH molecule and used \citet{Wende:2010iy} for
the line list, \citet{Sharp:2007kv} broadening coefficients, and
\citet{Visscher:2010iz} for the chemical equilibrium of 
FeH. In \citet{Tremblin:2015fx}, we applied the calculations to T dwarfs and showed that
a reduced temperature gradient in the atmosphere, as reproduced with
an effective adiabatic index ($\gamma \sim$~1.2 - 1.3) lower than the
equilibrium value ($\gamma \sim$~1.3 - 1.4) correctly reproduces the observed
spectra and the reddening in J - H colors, yielding a synthetic
spectrum similar to the one with the inclusion of clouds
\citep{Morley:2012io}.
We have explored these effects on the EGP
HR8799c (Fig.~\ref{fig:spec_hr8799c}) and the L dwarf Denis-P 0255
(Fig.~\ref{fig:spec_d0255}). As seen in 
figures \ref{fig:spec_hr8799c} and \ref{fig:spec_d0255}, the models
reproduce the observed spectra very well 
provided that (i) we include non-equilibrium chemistry and quench
CO/CH$_4$ (i.e. prevent the formation of CH$_4$), (ii) we decrease the
adiabatic index to $\gamma \sim$~1.05 in the lower part of their
atmospheres.
The only difference with the T-dwarf modeling of
\citet{Tremblin:2015fx} is 
that we modify the adiabatic index only within the layers corresponding to the emerging
flux bands.
The global effect of the modification of the adiabatic
index along the L sequence and the L/T transition is shown in Fig.~\ref{fig:spec_lt},
and the modified layers are indicated in the pressure/temperature (PT)
profiles in Fig.~\ref{fig:pt}.

The data for HR8799c are taken from \citet{Oppenheimer:2013gy},
\citet{Ingraham:2014gx}, and \citet{Skemer:2014hy}. In Fig.~\ref{fig:spec_hr8799c},
the blue spectrum is the out-of-equilibrium model obtained with a
value of the turbulent diffusion coefficient K$_{zz}$ = 10$^{11}$ cm$^2$s$^{-1}$. Such a
high value is consistent with the convective velocities computed in
  the model for these low gravities.
The 3.78-$\mu$m flux is higher in our model,
but PH$_3$ opacity could be significant around 4 $\mu$m  and is
not included in our models \citep[see][]{Morley:2014gs}.
Our out-of-equilibrium cloudless model with a modified adiabatic index
provides a better fit to the NIR data between 1 and 2 $\mu$m
compared to the cloud models \citep{Ingraham:2014gx}. The effective temperature, surface
gravity and radius we derive are compatible with evolutionary models
\citep{Baraffe:2003bj} and suggest a lower-mass, younger object (∼ 3
M$_\mathrm{Jup}$ at 10 M$_\mathrm{yrs}$) 
than previous studies. Since we used a higher metallicity, however,
different C/O ratios could also lead to a good fit with slightly
different values of effective temperature, surface gravity and
radius. These effects need to be investigated in more details in
follow-up studies.

The spectrum of Denis-P 0255 was obtained from \citet{Cushing:2005ed}. This object is a
typical late type L dwarf (L8) which has a characteristic reddening in
J - H and J - K. The model we used includes a zone with a modified
adiabatic index $\gamma$=~1.05 in the deep atmosphere. As a result, the
atmospheric profile has cooler deep layers and hotter upper ones than
the equilibrium atmosphere (see Fig.~\ref{fig:pt}). As a consequence, the fluxes
in Y and J bands are reduced and the ones in H and K bands are
enhanced. The modeled spectrum shows that this effect adequately
reproduces the main characteristics of the data, provided CH$_4$ is still
reduced compared to the equilibrium value (with now a value K$_{zz}$ =
10$^{7.5}$ cm$^2$s$^{-1}$, consistent, again, with the convective
velocities computed in the model
for these gravities). The effect is similar to what was inferred in
\citet{Tremblin:2015fx} for T dwarfs at the N$_2$/NH$_3$ transition. L dwarfs just require a
lower value of the adiabatic index in order to produce a stronger
reddening.

\section{L/T transition and FeH resurgence}\label{sect:feh}

From our T-dwarf modeling in \citet{Tremblin:2015fx}, it was unclear whether the need to
use cooler deep atmospheric layers was due to a real effect (fingering
convection) or was rather a way to mimic the effect of
clouds. The left panel of Fig.~\ref{fig:spec_lt} displays the evolution of the
spectrum along the L sequence (from 
T$_\mathrm{eff}$ = 1600 K to 1400 K). As expected, we clearly see a weakening of
the FeH spectral signature as a function of decreasing effective
temperature. The right panel of Fig.~\ref{fig:spec_lt} illustrates
the evolution of the spectrum at 
constant T$_\mathrm{eff}$ (= 1400 K) along the L/T transition when varying the
modified adiabatic index from a value $\gamma$ = 1.05, for L dwarfs, to $\gamma$ =
1.2-1.3, for early T dwarfs. Observations \citep{Golimowski:2004en} indeed suggest that the
L/T transition occurs at relatively constant effective
temperature. Figure~\ref{fig:spec_lt} clearly shows that along this transition our
models naturally yield a resurgence of the FeH absorption feature and
a brightening of the J band flux similar to what was found in
observations.

The L dwarf reddening is due to cooler deep atmospheric
layers than the ones corresponding to a radiatively stable atmosphere,
a consequence, as detailed in the next section, of the enhanced energy
transport due to local chemical turbulence, triggered by the
thermo-chemical instability. As one moves towards cooler atmospheres
when transitioning towards T dwarfs, the instability
vanishes. Turbulent dissipation warms up the deep atmosphere,
steepening the temperature gradient (yielding a larger effective
adiabatic index), increasing the FeH abundance and the flux in these
hotter layers, compared to the L-dwarf profile. This global effect is
illustrated in Fig.~\ref{fig:pt} . Along the L sequence, the atmospheric profiles
keep cooling down, notably in the deeper layers (P $\ge$ 1 bar), due to
the smaller thermal gradient (i.e. the smaller effective adiabatic
index). Along the L/T transition, the deep layers warm up, yielding a
steeper temperature gradient, while the upper layers keep cooling
down. An alternative suggested scenario to
explain the resurgence of FeH and the J band brightening is holes in
the cloud cover
\citep{Ackerman:2001gka,Burgasser:2002jq,Marley:2010kx} although it
appears to be partly excluded by \citet{Buenzli:2015tn}. 
We rather suggest that the proper explanation for 
these puzzling spectral behaviors is (i) cooler deep atmospheres for L
dwarfs than obtained with cloudy atmospheres, the conventional
explanation for the past 20 years, yielding the J - H and J - K
reddening, and (ii) a warming up of the deep layers at the L-T
transition, due to small-scale turbulent dissipation, which explains
both the FeH resurgence and the J band brightening.


\section{Instability of the CO/CH$_4$ and N$_2$/NH$_3$ transition}\label{sect:inst}

Although we have a model that can reproduce all the observed effects,
we need an explanation for the change in the modified adiabatic index
and the transition from $\gamma$ = 1.05 to 1.2-1.3. In \citet{Tremblin:2015fx}, we proposed that
small scale ''diffusive'' turbulence, more efficient than radiative
transport, induced by fingering convection triggered by thin dust
condensation, would be responsible for the decrease of the temperature
gradient. We revisit this scenario and rather suggest that the real
culprit is the instability of carbon and nitrogen chemistry in BD and
EGP atmospheres. Within the temperature range of interest, carbon is
preferentially in form of CO at high temperature and of CH$_4$ at low
temperature. Similarly, nitrogen changes from N$_2$ at high temperature
to NH$_3$ at low temperature. The net reactions for these transitions
are:
\begin{eqnarray}\label{eq}
  \mathrm{CO}+3\mathrm{H}_2 &\rightarrow& \mathrm{CH}_4+\mathrm{H}_2\mathrm{O}\cr
  \mathrm{N}_2 + 3\mathrm{H}_2 &\rightarrow & 2\mathrm{NH}_3  
\end{eqnarray}
It is quite clear that the part of the atmosphere in the methane or
ammonia dominant state will have a higher mean molecular weight than
the corresponding CO or N$_2$ dominated one, since there are globally
fewer molecules in these latter states. On the other hand, it is well
known from chemical network studies that the chemistry for these
transitions can be very slow \citep{Venot:2012fr,Moses:2011bn}. Therefore, atmospheres with
these chemical transitions can develop an instability similar in
nature to fingering double diffusive convection, but with the chemical
reactions themselves now playing the role played otherwise by
molecular diffusion, leading to a thermo-chemical instability. The
analogy is as follows: at the CH$_4$/CO transition, if a perturbation
drives some ''CH$_4$+H$_2$O'' mixture down in the ''CO+3H$_2$'' (warmer) deeper
layers, and if we are in the stable radiative part of the atmosphere,
because of the slowness of the chemical reaction the mixture will stay
in its methane-rich state and sink because of its higher molecular
weight. The same process can happen at the N$_2$/NH$_3$ transition. The
extent of the unstable zone can be estimated by using an extension of
the Ledoux criterion \citep{Rosenblum:2011jb}:
\begin{equation}\label{eq:ledoux}
 R_0 = \frac{\nabla_T-\nabla_\mathrm{ad}}{\nabla_\mu} <\frac{1}{\tau}= \frac{\kappa_T}{l^2/\tau_\mathrm{chem}}
\end{equation}

\begin{figure}[t]
\centering
\includegraphics[width=\linewidth]{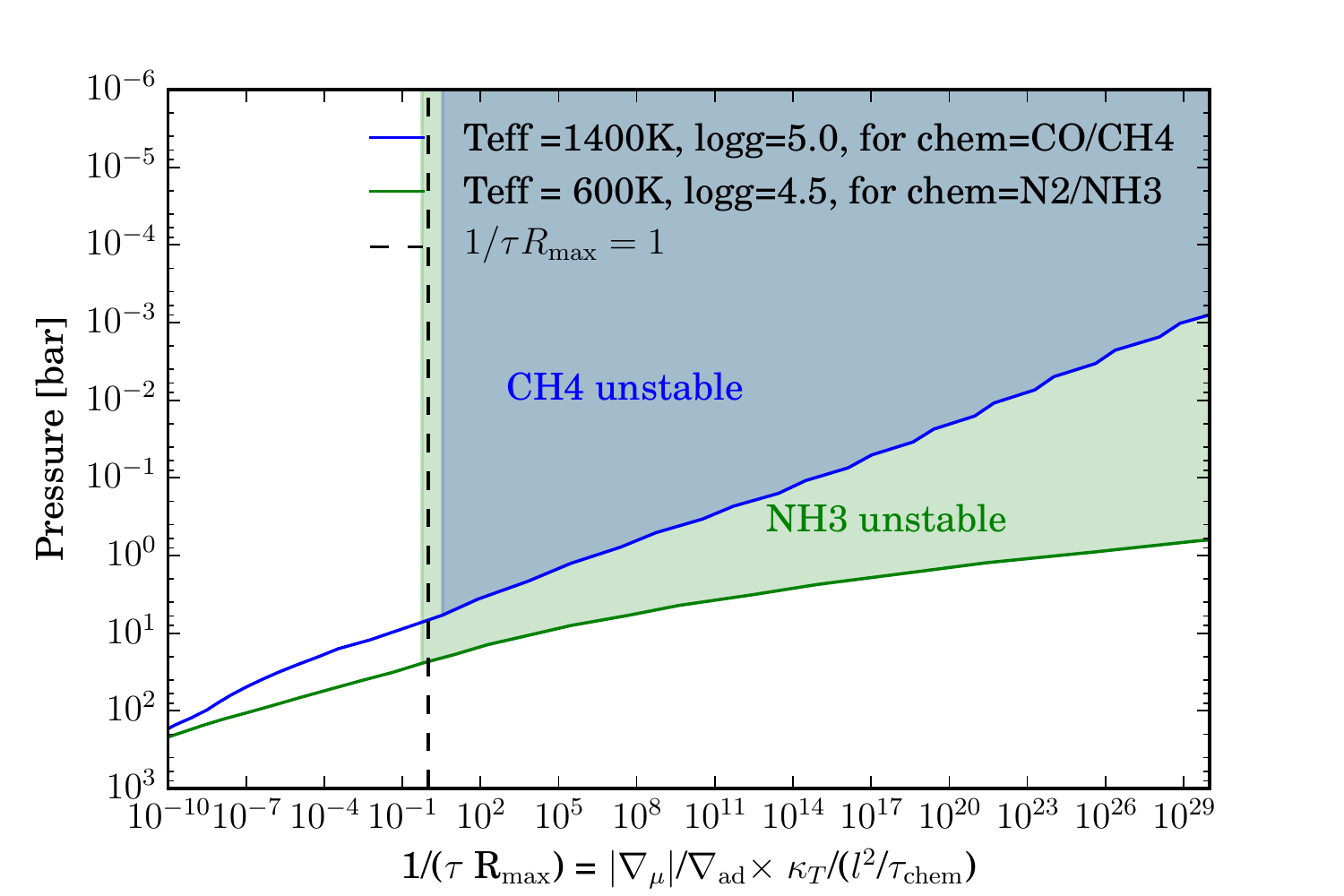}
\includegraphics[width=\linewidth]{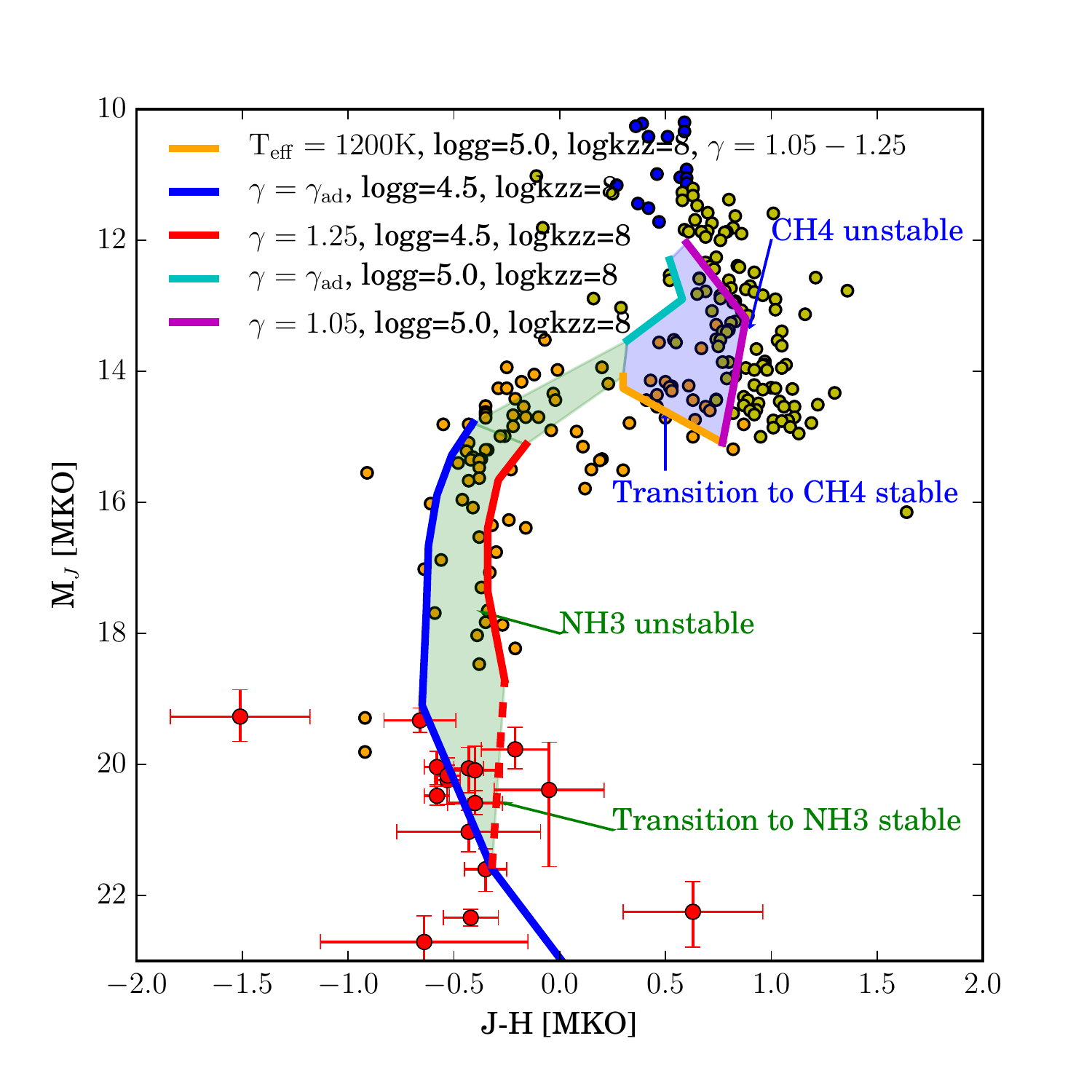}
\caption{\label{fig:chem_cmd} Top: Profiles of 1/($\tau$ R$_\mathrm{max}$) for
  a typical L dwarfs with a CO/CH$_4$ transition and for a typical T
  dwarf with a N$_2$/NH$_3$ transition. The part of the atmosphere unstable to
  the modified Ledoux criterion is indicated with the shaded
  colors. Bottom: M$_J$ versus $J-H$ color magnitude diagram
  for brown dwarfs compared to our equilibrium models (blue and cyan)
  and our out-of-equilibrium adiabatic-index-modified models (red and
  magenta). We also plot (as a guideline) the transition from L to T at
    a constant effective temperature of 1200 Kelvin (orange). We
  indicate the possible zone where the objects will be 
  subject to the thermo-chemical instability if the CO/CH$_4$
  transition and the N$_2$/NH$_3$ transition. The Y-dwarf
  photometry is from \citet{Dupuy:2013ks,Beichman:2014jr} and the
  L/T/M from \citet{Dupuy:2012bp,Faherty:2012cy}.}
\end{figure}

We have replaced the usual molecular diffusion coefficient by $l^2/\tau_\mathrm{chem}$ 
with $\tau_\mathrm{chem}$ the timescale of the chemical reaction and $l$ the typical
size of the unstable structures. We have computed the timescales for
the reactions using the equations given by \citet{Zahnle:2014hl}, and we assumed that
the typical size of the unstable structures is 1\% of the scale height
\citep[see][for a
  discussion on the size of the fingers expected in the stellar
  context]{Traxler:2011dx}. In order to be conservative, we use an upper limit
for $R_0$, given by
$R_\mathrm{max}=\nabla_\mathrm{ad}/|\nabla_{\mu}|$. The unstable
regions are portrayed 
in the top panel of Fig.~\ref{fig:chem_cmd} for the CO/CH$_4$ chemistry in an L dwarf at
T$_\mathrm{eff}$  = 1400 K and 
for the N$_2$/NH$_3$ chemistry for a T dwarf at T$_\mathrm{eff}$  = 600 K. For
simplicity we derived the mean-molecular-weight gradient $\nabla_\mu$ from the difference between a full ammonia or
methane state and a full CO or N$_2$ state over one scale height $H$,
reduced by a factor $l/H$ (i.e. $\nabla_\mu \sim$~-10$^{-5}$ for
CO/CH$_4$ and $\nabla_\mu \sim$~-10$^{-6}$ for
N$_2$/NH$_3$). The left panel of Fig.~\ref{fig:chem_cmd} shows that the chemical transitions in all the upper
atmosphere of a L or T dwarf are unstable, because of the slow
chemistry. 
In the bottom panel of Fig.~\ref{fig:chem_cmd}, we compare the color magnitude diagram (CMD) M$_J$ versus J -
H obtained with the equilibrium models and with the models with
out-of-equilibrium chemistry and a modified adiabatic index. As for
the T dwarfs in \citet{Tremblin:2015fx}, the L-dwarf reddening is well reproduced in the
CMD with the temperature-gradient reduction. Other CMDs with the NIR
bands Y, J, H, and K give a similar good match between our models and
the observations since, as shown in Fig.~\ref{fig:spec_d0255}, the NIR spectra are well
reproduced. We also illustrate the transition from L to T with
  a model at constant effective temperature and an adiabatic index
  varying from 1.05 (L) to 1.25 (T) (orange line in Fig.~5). This highlights the occurence of the J-band
brightening in our (cloudless) 1D models, even though a detailed
sequence based on consistent evolutionary models is needed to properly take into account
the evolution of gravity, effective temperature and modified
adiabatic index. This will be explored in a forthcoming paper.
We indicate the domains where the CO/CH$_4$ and N$_2$/NH$_3$
transitions are predicted to be unstable. The sharpness of the limits
corresponding to the instability, according to the above modified
Ledoux criterion, naturally explains the sharpness of the L/T
transition. Indeed, as temperature decreases, reaching the later T
dwarf regime, the CO/CH$_4$ transition will lie at deeper levels than the
ones subject to the instability, so that the now CH$_4$-dominated
atmosphere will become stable again, as the chemical reaction timescale $\tau_\mathrm{chem}$ becomes shorter than the radiative timescale. This does not preclude, however,
the presence of some (quenched) CH$_4$ in the upper layers of L dwarf
atmospheres. 

\section{Discussion and conclusions} \label{sect:disc}

In this paper, we show that:
\begin{itemize}
\item Conditions characteristic of the atmospheres of L dwarfs favor
  the onset of a thermo-chemical instability at the CO/CH$_4$ transition,
  while conditions typical of T dwarf atmospheres yield the same
  instability at the N$_2$/NH$_3$ transition. The instability is due to the
  development of a destabilizing molecular weight gradient induced by
  the slowness of these chemical equilibrium reactions. This gradient
  generates local chemical (compositional) convection, which decreases
  the temperature gradient in the atmosphere.
\item The spectra of L dwarfs and hot EGPs is well reproduced if (i)
  the temperature gradient in the atmosphere is decreased, because of
  the aforementioned instability, and (ii) if the quenching of CH$_4$ at
  the CO/CH$_4$ transition is taken into account. The same picture
  applies to T dwarfs, with quenching of NH$_3$ at the N$_2$/NH$_3$
  transition. 
\item As the instability vanishes along the L/T transition, small
  scale turbulent dissipation warms up the deep layers and thus
  increases the temperature gradient in the atmosphere. This naturally
  explains the so far unexplained FeH resurgence and J-band
  brightening in early T dwarfs.
\end{itemize}

A detailed numerical analysis of the instability is certainly needed
to determine the growth rate and the size of the most unstable mode
and thus the typical time and space scales of the instability. The
efficiency of the induced small-scale turbulent energy transport and
dissipation also need to be determined in order to properly quantify
(i) the modified adiabatic index, (ii) the amount of turbulent heating
which yields an increase of temperature in the deep atmospheric layers
at the L/T transition. Such studies require non-trivial numerical
explorations.

Although by no means providing a proof, several features bring support
to the reality of this instability. First, its disappearance when the
atmosphere becomes stable again (according to condition \ref{eq:ledoux}) could
explain the sharpness in effective temperature of the L/T
transition. Second, turbulent energy transport, triggered by the
instability of the CO/CH$_4$ and N$_2$/NH$_3$ transitions, decreases the
temperature gradient in the atmosphere, leading to the reddening of L
and T dwarfs, respectively. Third, the strength of the instability
being intrinsically linked to the magnitude of the
mean-molecular-weight gradient, the fact that the gradient associated
to the CO/CH$_4$ transition is larger than the one at the N$_2$/NH$_3$
transition \citep[because of the C/N $\sim$ 4 abundance
  ratio][]{Asplund:2009eu} explains 
the stronger reddening for L dwarfs than for T dwarfs. Fourth, the
turbulence induced by CO or temperature fluctuations in the atmosphere
during the transition from CO to CH4 at the L/T transition could
explain the observed variability. Thus the fluctuations observed by
Doppler imaging reflect the ones in CO abundances or in temperature
\citep{Crossfield:2014cy}.

Given the coherence of this global picture, we conclude that clouds
are not needed to explain the main characteristics of the emission
spectra of BDs and directly imaged EGPs. We emphasize that this does
not imply an absence of clouds. Given the possibility of many stable
condensates, and the presence, according to the present scenario, of
turbulent layers, these atmospheres very likely present some cloud
cycles. Furthermore, some observations, such as the absorption in the 10~$\mu$m
window \citep[e.g.][]{Cushing:2006hx}, suggest the presence of condensates.
 Current cloud models, however,
do not well reproduce this signature \citep[see][]{Stephens:2009cc}, requiring a revision
of the models in
order to reproduce both this signature and the NIR color reddening.

 Not only
does the present scenario completely change our present understanding
(and non understanding) of BD and EGP atmospheres, but, if correct, it
shows that the same physical mechanism, namely chemical or fingering
convection, induced by a thermo-chemical instability, would take place
in environments as different as Earth oceans \citep{Rahmstorf:2003vg}, the Earth core-mantle
boundary \citep{Hansen:1988in}, and exoplanet and brown dwarf atmospheres, nicely
illustrating the universality of physical processes in nature.

%
%

\begin{acknowledgements}
We thank Patrick Ingraham and Rebecca Oppenheimer for providing their
data. This work is partly supported by
the European Research Council under the European Community's Seventh
Framework Programme (FP7/2007-2013 Grant Agreement No. 247060 and
FP7/2007-2013 Grant Agreement No. 247060-PEPS and grant 
No. 320478-TOFU).  
Part of this work is supported by the Royal Society award
WM090065. O.V. acknowledges 
support from the KU Leuven IDO project IDO/10/2013 and from the FWO
Postdoctoral Fellowship programme. 
\end{acknowledgements}

%
%



\end {document}